\documentclass[12pt,reqno]{article}
\usepackage{amsfonts}
\usepackage{amsmath}
\usepackage{amsbsy}

\usepackage{amssymb,latexsym}
\numberwithin{equation}{section}

\begin{document}

\title{Exact Solutions in Five-Dimensional Axi-dilaton Gravity with Euler-Poincar\'{e} Term}
\begin{titlepage}
\author{ A. N. Aliev\footnote{E.mail: aliev@gursey.gov.tr} \\
{\small Feza G\"{u}rsey Institute, 34684 \c{C}engelk\"{o}y,
\.{I}stanbul, Turkey} \\ \\
H. Cebeci\footnote{E.mail:
hcebeci@anadolu.edu.tr} \\
{\small Department of Physics, Anadolu University, 26470 Eski\c{s}ehir, Turkey} \\ \\
T. Dereli\footnote{E.mail: tdereli@ku.edu.tr}\\{\small Department
of Physics,  Ko\c{c} University, 34450 Sar{\i}yer-\.{I}stanbul,
Turkey} }

\date{ }

\maketitle

\bigskip

\begin{abstract}

\noindent We examine the effective field equations that are
obtained from the axi-dilaton gravity action with a second order
Euler-Poincar\'{e} term and a cosmological constant in all higher
dimensions. We solve these equations for five-dimensional
spacetimes possessing homogeneity and isotropy in their
three-dimensional subspaces. For a number of interesting special
cases we show that the solutions  fall into two main classes: The
first class consists of time-dependent solutions with spherical or
hyperboloidal symmetry which require certain fine-tuning relations
between the coupling constants of the model and the cosmological
constant. Solutions in the second class are locally static and
prove the validity of Birkhoff's staticity theorem in the
axi-dilaton gravity. We also give a  class of static solutions,
among them  the well-known charged black hole solutions with a
cosmological constant in which the usual electric charge is
superseded by an axion charge.

\end{abstract}

\bigskip

\end{titlepage}

\section{Introduction}

Developments in string/M-theory have led to a revolutionary idea
that our universe is  a slice, a {\it braneworld}, in a
higher-dimensional spacetime \cite{horava}-\cite{lukas}. The most
striking  phenomenological consequences of the braneworld idea
have been studied in Large-Extra-Dimension scenarios
\cite{ADD,RS}. These scenarios open up the possibility of an
elegant geometric resolution of the large hierarchy between the
electroweak scale and the fundamental scale of quantum gravity as
well as of directly probing TeV-scale mini black holes in
high-energy collisions \cite{gt}. They also support the properties
of four-dimensional Einstein gravity in low energy limit that is
verified by examining the effective gravitational field equations
in the braneworld.  The gravitational field equations on a $ Z_{2}
$-symmetric 3-brane in a five-dimensional bulk spacetime were
studied in \cite{sms1, aemir1}. Some exact solutions to these
equations that describe black holes localized on the 3-brane were
found in \cite{dmpr, aemir2}.

Cosmological dynamics of the braneworld universe in general
exhibits significant deviation from standard
Friedmann-Robertson-Walker picture. However, it appears very
similar to the latter at late times. This has been examined within
two basic approaches: In the first approach our universe is
supposed to be a fixed brane in a time-evolving bulk spacetime
\cite{david}, while in the second one it is a moving brane in a
static bulk spacetime \cite{kraus}. It is remarkable that the most
general static AdS solution in five dimensions that  induces the
FRW cosmology on the brane is the Schwarzschild-AdS solution with
constant 3-space curvature \cite{ida, bcr}.

The results described above have also provided a strong impetus
for examining the braneworld idea in string-generated gravity
models. Of these, the models with quadratic-curvature correction
term (Gauss-Bonnet combination) are of particular interest
\cite{zwiebach}. The Gauss-Bonnet term is the leading order
correction to the ordinary Einstein-Hilbert action in the
low-energy limit of string theory. It is also sometimes called a
second order Euler-Poincar\'{e} density term. The static solutions
for black holes with asymptotic AdS behavior in these models were
found in \cite{boulware}-\cite{rong}(for a static wormhole
solution see recent works in Refs. \cite{ricardo, gravanis}),
while cosmological dynamics on the 3-brane has been studied in
\cite{germani}-\cite{lidsey} (see also a review paper \cite{roy}
and references therein). The effective gravitational field
equations on the 3-brane imprinted by a five-dimensional
Einstein-Hilbert action with a second order Euler-Poincar\'{e}
density term were obtained in \cite{mt, acd}.

In light of all this, it becomes clear that further study of exact
solutions in string-generated gravity models is of particular
interest. Here we shall study the exact static solutions in
five-dimensional axi-dilaton gravity with a second-order
Euler-Poincar\'{e}  term,  focusing on the bulk theory. As is
known, the axi-dilaton gravity models naturally arise in the
low-energy limit of superstring theories. The corresponding
effective actions are determined by an expansion in a small string
tension parameter $ \alpha^{\prime} $ and, in addition to the
usual Einstein-Hilbert term with a cosmological constant, they
also include a dilaton $ 0 $-form, a Yang-Mills $ 1 $-form, an
axion $ 3 $-form and second and higher-order Euler-Poincar\'{e}
densities \cite{dereli1}-\cite{lovelock}.

The paper is organized as follows. In Sec. II, using the formalism
of differential forms, we present the action of our axi-dilaton
gravity model in the Einstein frame and in all higher dimensions.
To simplify the model we do not include the Yang-Mills coupling in
the action. The effective field equations are obtained by varying
this action with respect to all dynamical variables of the model.
In order to avoid the appearance of propagating torsion in the
variational procedure \cite{dereli1}, we  constrain the space of
the dynamical variables to be torsion free. In Sec. III we
restrict ourselves to five-dimensional spacetimes and solve the
effective field equations for the spacetimes possessing
homogeneity and isotropy in their three-dimensional subspaces. We
examine a number of interesting special cases and show that the
solutions  fall into two classes: The first class consists of
time-dependent solutions with spherical or hyperboloidal symmetry
which require certain fine-tuning relations between the coupling
constants of the model and the cosmological constant. Solutions in
the second class are locally static and prove the validity of
Birkhoff's staticity theorem in the axi-dilaton gravity. Finally,
in Sec. IV we give  some explicit  static solutions in five
dimensions, among them the well-known  black hole solution in
which the usual electric charge is superseded  by an axion charge.

\section{Action}

We consider a $ d $-dimensional manifold $ M $ endowed with a
spacetime metric $ {\mbox{\boldmath${g}$}} $ and examine the
action of axi-dilaton gravity in this spacetime. In
\cite{dereli1}, it has been shown that if all fields in the action
are taken to be unconstrained, then the corresponding variational
procedure inevitably yields propagating torsion. To circumvent it,
we constrain the connection $ 1 $-forms to be torsion-free. In
addition to the Einstein-Hilbert term with a cosmological
constant, the action includes a dilaton $ 0 $-form field $ \phi $,
an axion $ 3 $-form field $ H  $ and a second-order
Euler-Poincar\'{e} term. In the spirit of a heterotic string
model, we also impose an anomaly-freedom constraint or a Bianchi
condition on the axion field. The $ d $-dimensional action has the
form
\begin{equation}
I[e, \omega, \phi, H] = \int_M {\cal L}, \label{1}
\end{equation}
where in the Einstein frame the Lagrangian density $ d $-form is
given by \cite{dereli1}
\begin{eqnarray}
{\cal L} &=& \frac{1}{2}\, R^{ab} \wedge \ast (e_{a} \wedge e_{b}
) - \frac{\alpha}{2}\, d \phi \wedge \ast d \phi +
\frac{\beta}{2}\, e^{ - \beta_{2} \phi} H \wedge \ast H + \Lambda \,e^{ - \beta_{1} \phi} \ast 1  \nonumber
\\[2mm]
& & + \,\frac{\eta}{4}\, R^{ab} \wedge R^{cd} \wedge \ast ( e_{a}
\wedge e_{b} \wedge e_{c} \wedge e_{d} )   \nonumber \\[2mm]
 & & + \,( d e^{a} + \omega^{a}\,_{b} \wedge e^{b} ) \wedge \lambda_{a} + ( d H -
\frac{ \varepsilon }{2} \,R_{ab} \wedge R^{ab} ) \wedge \mu \; .
\label{2}
\end{eqnarray}
Here the basic gravitational field variables are the co-frame
1-forms $ e^{a} $ in terms of which, the spacetime metric is
decomposed as  $ {\mbox{\boldmath${g}$}}=\eta_{ab}\, e^{a} \otimes
e^{b}$ with $\eta_{ab} = diag ( - + + + \cdots )\,$. The free
parameters $ \alpha \,$, $ \beta \,$, $\eta \,$, $ \beta_1 $, $
\beta_2 $ and $ \varepsilon $ are the corresponding coupling
constants. The star $\ast$ denotes the Hodge dual map and $
\Lambda $ is the cosmological constant. The quantities $
\lambda_{a} $ and $ \mu $ are the Lagrange multiplier forms that
are used to impose the torsion-free and anomaly-freedom
constraints in the variational procedure \cite{dereli1}.

The torsion-free, Levi-Civita connection $ 1 $-forms $\,
\omega^{a}\,_{b} \,$ obey the first Cartan structure equations
\begin{equation}
d e^{a} + \omega^{a}\,_{b} \wedge  e^{b} = 0 \label{3}
\end{equation}
where  $\omega_{a \,b} = - \omega_{b \,a}\, $, while the
corresponding curvature 2-forms  follow from the second Cartan
structure equations
\begin{equation}
R^{a b} = d \omega^{a b} + \omega^{a}\,_{c} \wedge
\omega^{c\,b}\,\, . \label{4}
\end{equation}
The second order Euler-Poincar\'{e} form \cite{dereli2, dereli3}
is given by
\begin{equation}
{\cal L}_{EP} =  R^{a b} \wedge R^{c d} \wedge \ast ( e_{a} \wedge
e_{b} \wedge e_{c} \wedge e_{d} )\,\,, \label{5}
\end{equation}
which can also be written in the alternative form
\begin{equation}
{\cal L}_{EP} = 2 R_{a b} \wedge \ast R^{a b} - 4 P_{a} \wedge
\ast P^{a} + {\cal R}_{(d)}^{2} \ast 1 \label{6}\,\,,
\end{equation}
where we have used the  Ricci 1-form $ P^{a} = \iota_{b} R^{b a} $
and the curvature scalar $ {\cal R}_{(d)} = \iota_{a}\,\iota_{b}\,
R^{ba}  $\,\,. Here $ \iota_{a}= \iota_{X_a} $ is the interior
product operator such that $ \iota_{X_a} (e^b) = \delta^{a}_b $.

\subsection{Field equations}

The effective field equations are obtained from independent
variations of the total action density in (\ref{2}) with respect
to the variables $ e^{a} $,  $ \omega^{a}\,_{b} \,$, $ \phi $ and
$ H $. The variations with respect to co-frame fields $ e^{a} $
yield the Einstein equations
\begin{eqnarray}
\frac{1}{2} \,R^{ab} \wedge \ast ( e_{a} \wedge e_{b} \wedge e_{c}
) = - \frac{\alpha}{2}\, \tau_{c} [ \phi ] +
\frac{\beta}{2}\, e^{ - \beta_{2} \phi} \tau_{c} [H] - \Lambda\, e^{ - \beta_{1} \phi} \ast e_{c} \nonumber
\\[2mm]
- \frac{\eta}{4}\, R^{ab} \wedge R^{f g} \wedge \ast ( e_{a}
\wedge e_{b} \wedge e_{f} \wedge e_{g} \wedge e_{c} ) - D
\lambda_{c} \,\,, \label{7}
\end{eqnarray}
where
\begin{equation}
\tau_{c} [\phi] = \iota_{c} d \phi \wedge \ast d \phi + d \phi
\wedge \iota_{c} ( \ast d \phi )\,\, \label{8}
\end{equation}
and
\begin{equation}
\tau_{c} [H] = \iota_{c} H \wedge \ast H + H \wedge \iota_{c} (
\ast H) \,\,\label{9}
\end{equation}
are the stress-energy $(d-1)$-forms for the dilaton and axion
fields, respectively. Here and in what follows $ D $ denotes the
exterior covariant derivative operator. From variations with respect
to $\phi$, we obtain the dilaton field  equation
\begin{equation}
\alpha d ( \ast d \phi ) = \frac{\beta_{2} \beta}{2} \, e^{ -
\beta_{2} \phi} H \wedge \ast H + \Lambda \beta_{1}  e^{ -
\beta_{1} \phi} \ast 1 \,\,. \label{10}
\end{equation}
From the connection $\omega^{a}\,_{b}$ variations we obtain
\begin{eqnarray}
\frac{1}{2}\, ( e^{a} \wedge \lambda^{b} - e^{b} \wedge
\lambda^{a} ) &=& \frac{1}{2}\, D \left( \ast ( e^{a} \wedge e^{b}
) \right)
- \varepsilon \,D ( R^{ab} \wedge \mu ) \nonumber \\
& & + \,\frac{1}{2} \, \eta D \left( R^{cd} \wedge \ast ( e^{a}
\wedge e^{b} \wedge e_{c} \wedge e_{d} ) \right) . \label{11}
\end{eqnarray}
Finally, the variations  with respect to $ H $ give the equation
\begin{equation}
d \mu = - \beta e^{ - \beta_{2} \phi } \ast H . \label{12}
\end{equation}
By taking the exterior derivative of both sides of this equation
we obtain the axion field equation
\begin{equation}
d ( e^{ - \beta_{2} \phi } \ast H ) = 0 \,\,, \label{13}
\end{equation}
since $  d^2 \mu =0 $. We also have the equation
\begin{equation}
d H = \frac{ \varepsilon }{2} \,R_{ab} \wedge R^{ab} \,\,.
\label{13'}
\end{equation}
Using the identity
\begin{equation}
D R^{ab} = 0  \label{14}
\end{equation}
one can solve equation (\ref{11}) uniquely for the Lagrange
multipliers $ \lambda^{a} $; we have
\begin{eqnarray}
\lambda^{a} &=& 2 \,\varepsilon \beta\, e^{ - \beta_{2} \phi}
\iota_{b} ( R^{ba} \wedge \ast H ) - \frac{\varepsilon \beta
}{2}\, e^{ - \beta_{2} \phi } \,e^{a} \wedge \iota_{s} \iota_{l} (
R^{l\,s} \wedge \ast H ) \,\,. \label{15}
\end{eqnarray}
Next, substituting this expression into the Einstein field
equations in (\ref{7}) we put them in the form
\begin{eqnarray}
\frac{1}{2}\, R^{ab} \wedge \ast ( e_{a} \wedge e_{b} \wedge e_{c}
) = - \frac{ \alpha }{2} \, \tau_{c} [\phi] +
\frac{\beta}{2}\, e^{ - \beta_{2} \phi} \tau_{c} [H] - \Lambda \,e^{ - \beta_{1} \phi} \ast e_{c} \nonumber
\\[2mm]
- \frac{\eta}{4} \,R^{ab} \wedge R^{f g} \wedge \ast ( e_{a}
\wedge e_{b} \wedge e_{f} \wedge e_{g} \wedge e_{c} ) \nonumber
\\[2mm]
- 2 \varepsilon \beta D ( e^{ - \beta_{2} \phi} \iota_{b} (
R^{b}\,_{c} \wedge \ast H ) ) - \frac{\varepsilon \beta}{2}\,
e_{c} \wedge D ( e^{ - \beta_{2} \phi} \iota_{s} \iota_{l} (
R^{l\,s} \wedge \ast H ) ) \,\,, \label{16}
\end{eqnarray}
where the last two terms  can be thought of as non-trivial
improvement terms that appear due to the torsion-free and
anomaly-freedom constraints.

\section{Metric ansatz}

In this section we restrict ourselves to five-dimensional spacetimes
and seek to solve the effective field equations obtained above. It
is useful to start with a metric ansatz for a spacetime possessing
homogeneity and isotropy in its three-dimensional subspace. The most
general metric for this spacetime can be taken of the form
\begin{equation}
{\mbox{\boldmath${g}$}} = - f^{2} ( t, w ) \,d t^{2} + u^{2} ( t,w
)\, d w^{2} + g^{2} ( t,w )\,\, \frac{d x^{2} + d y^{2} + d z^{2}
}{\left( 1 +\frac{\kappa \,r^2}{4} \right)^{2}}\,\,,\label{17}
\end{equation}
where $ f $, $ u $ and $ g $ are arbitrary functions of the time
coordinate $ t $  and  the coordinate $\, w \,$ of the fifth
dimension. The curvature index $ \kappa= -1,\,0\,, + 1 $
corresponds to a hyperbolic, planar and spherical geometry of the
three-dimensional subspace, respectively and $ r^{2} = x^{2} +
y^{2} + z^{2}  $. We assume the same functional dependence for the
dilaton field
\begin{equation}
\phi = \phi \,( t,w )\,\,, \label{18}
\end{equation}
and for the axion $3$-form
\begin{equation}
H = h ( t, w )\, \frac{d x \wedge d y \wedge d z}{\left( 1 +
\frac{\kappa \,r^{2} }{4} \right)^{3}}\,\,\,. \label{19}
\end{equation}
The co-frame 1-forms  for the metric (\ref{17}) can be chosen as
\begin{equation}
e^{0} = f ( t,w )\, d t\,,~~~ e^{5} = u ( t,w )\, d w\,,~~~ e^{i}
= g ( t,w ) \frac{d x ^{i}}{\left( 1+ \frac{\kappa \,r^{2}}{4}
\right)}\,,~~~ x^i = x , y , z  . \label{20}
\end{equation}
Evaluating with these co-frame 1-forms the Levi-Civita connection,
we obtain
\begin{eqnarray}
\omega^{0}\,_{i} & = & \frac{g_{,\,t}}{f g}\, e^{i}\,\,,
~~~~~~~\omega^{i}\,_{j}  = \frac{\kappa}{2 g} \,( x^{i} e^{j} -
x^{j} e^{i}
)\,\,,\\[2mm]
\omega^{0}\,_{5} &= & \frac{u_{,\,t}}{f u}\, \,e^{5} +
\frac{f_{,\,w}}{f u}\,\, e^{0}\,\,, \qquad \omega^{i}\,_{5} =
\frac{g_{,\,w}}{u g}\,\, e^{i}\,\, , \label{22}
\end{eqnarray}
where commas denote partial derivatives with respect to an
appropriate index. For the corresponding curvature 2-forms we find
the expressions
\begin{eqnarray}
R^{05}&=& B \,e^{0} \wedge e^{5}\,\,, \qquad R^{ij} = A\, e^{i}
\wedge
e^{j}\,\,,  \\[2mm] R^{0 i} & = & C \,e^{0} \wedge e^{i} + D\, e^{5} \wedge
e^{i}\,\,,\\[2mm]
R^{i 5}& = & D \,e^{0} \wedge e^{i} + E\, e^{5} \wedge e^{i}\,\,,
\label{25}
\end{eqnarray}
where
\begin{equation}
A = \frac{1}{g^{2}} \left[ \kappa + \left( \frac{g_{,\,t}}{f}
\right)^{2} - \left( \frac{g_{,\,w}}{u} \right)^{2} \right]\,\, ,
\label{26}
\end{equation}
\begin{equation}
B = \frac{1}{f u} \left[ \left( \frac{u_{,\,t}}{f} \right)_{,\,t}
- \left( \frac{f_{,\,w}}{u} \right)_{,\,w} \right]\,\, ,
\label{27}
\end{equation}
\begin{equation}
C = \frac{1}{f g} \left[ \left( \frac{g_{,\,t}}{f} \right)_{,\,t}
- \frac{f_{,\,w}\, g_{,\,w} }{u^{2}} \right] \,\,, \label{28}
\end{equation}
\begin{equation}
D = \frac{1}{ug} \left[ \left( \frac{g_{,\,t}}{f} \right)_{,\,w} -
\frac{ u_{,\,t}\, g_{,\,w} }{f u} \right]\,\, , \label{29}
\end{equation}
\begin{equation}
E = \frac{1}{ug} \left[ \left( \frac{g_{,\,w}}{u} \right)_{,\,w} -
\frac{g_{,\,t}\, u_{,\,t}}{f^{2}} \right]\,\, . \label{30}
\end{equation}
Using the above equations, it is straightforward to show that $
R_{ab} \wedge R^{ab} = 0 $. Then with ansatz (\ref{19}), equation
(\ref{13'}) yields
\begin{equation}
H = \frac{Q}{g^{3}} \,e^{1} \wedge e^{2} \wedge e^{3}\,\,,
\label{31}
\end{equation}
where the parameter $ Q $ can be thought of as  a charge
associated with the axion $3$-form field. Taking all this into
account in equations (\ref{10}) and (\ref{16}), we obtain the
following system of coupled partial differential equations
\begin{eqnarray}
 A + B + 2\, C - 2\, E = - 2 \eta \,\left( A\, B + 2\,
D^{2} - 2\, C\, E \right) \nonumber
\\[1mm]  - \frac{\alpha}{2} \,\left[ \left(\frac{\phi_{,\,t}}{f} \right)^{2} -
\left( \frac{\phi_{,\,w}}{u} \right)^{2} \right] +
\frac{\beta}{2}\, \frac{Q^{2}}{g^{6}}\, e^{ - \beta_{2}  \phi} -
\Lambda \,e^{ - \beta_{1} \phi } \,\,, \label{hc}
\end{eqnarray}
\begin{eqnarray}
3 A + 3 \,C(1+ 2\,\eta A) &=& - \frac{ \alpha }{2}\, \left[ \left(
\frac{ \phi_{,\,t} }{f} \right)^{2} + \left( \frac{\phi_{,\,w}}{u}
\right)^{2} \right] \nonumber \\[1mm]
& & - \frac{\beta}{2}\, \frac{Q^{2}}{g^{6}}\,\, e^{ - \beta_{2}
\phi} - \Lambda \,e^{ - \beta_{1}  \phi}\,\, , \label{33}
\end{eqnarray}
\begin{eqnarray}
3 A - 3  \,E(1+ 2\,\eta A) &=& \frac{ \alpha }{2} \,\left[ \left(
\frac{\phi_{,\,t}}{f} \right)^{2} + \left( \frac{\phi_{,\,w}}{u}
\right)^{2} \right] \nonumber \\[1mm]
 & & - \frac{\beta}{2}\, \frac{Q^{2}}{g^{6}}\,\, e^{ - \beta_{2}  \phi} - \Lambda \,e^{ - \beta_{1}  \phi} \,\,,
\label{34}
\end{eqnarray}
\vspace{1mm}
\begin{equation} \alpha \,\left[ \left( \frac{
\phi_{,\,w} \,g^{3} f}{u} \right)_{,\,w} - \left( \frac{
\phi_{,\,t}\, g^{3} u}{f} \right)_{,\,t}\right] \frac{ 1 }{g^{3} f
u} = \frac{\beta_{2} \beta}{2}\,\frac{Q^{2}}{g^{6}}\, e^{ -
\beta_{2}  \phi} + \Lambda \beta_{1} \,e^{ - \beta_{1}  \phi}
\,\,, \label{35}
\end{equation}
\vspace{1mm}
\begin{equation}
D ( 1 + 2 \,\eta A ) = - \frac{ \alpha }{3}\,
\frac{\phi_{,\,t}}{f}\, \frac{\phi_{,\,w}}{u}\,\,. \label{36}
\end{equation}
Next, we discuss the validity of Birkhoff's staticity theorem in
our model without invoking explicit solutions of this system of
equations.

\subsection{Birkhoff's theorem}

Following the similar procedure used in \cite{charm}, we combine
equations (\ref{33}) and (\ref{34}) to eliminate the cosmological
constant and  axion charge. This yields
\begin{equation}
(C + E)( 1 + 2 \,\eta  A ) = - \frac{ \alpha }{3} \left[ \left(
\frac{\phi_{,\,t} }{f} \right)^{2} + \left( \frac{ \phi_{,\, w }
}{u} \right)^{2} \right] \label{37}\,\, ,
\end{equation}
which  after combining with equation (\ref{36}) gives
\begin{equation}
( C + E \mp  2\, D ) ( 1 + 2\, \eta A ) = - \frac{ \alpha }{3}
\left[ \left( \frac{\phi_{,\,t}}{f} \right)^{2} + \left(
\frac{\phi_{,\,w}}{u} \right)^{2}  \mp \,2 \frac{\phi_{,\,t}}{f}
\frac{\phi_{,\,w}}{u} \right] \,\,. \label{39}
\end{equation}
With a constant  dilaton field  these equations serve as the
integrability conditions  for the field equations
(\ref{hc})-(\ref{36}). To facilitate further considerations we use
the conformal properties of the spacetime metric in $ (t,
w)$-plane. This allows us, without loss of generality, to take $ f
= u $ and pass to the lightcone coordinates
\begin{equation}
n = \frac{ t+ w }{2} , \qquad s = \frac{t-w}{2} \,\,. \label{40}
\end{equation}
Rewriting now equation (\ref{39}) in these coordinates, we arrive
at the following equations
\begin{equation}
\left( g_{,\,ss} - 2 \,\frac{f_{,\,s}}{f}\, g_{,\,s} \right)
\left[ g^{2} f^{2} + 2 \,\eta \left( \kappa f^{2} + g_{,\,n}
\,g_{,\,s} \right) \right] \frac{1}{f^{2} g^{3}}  = -
\frac{\alpha}{3}\, (\phi_{,\,s})^{2} \label{41}\,\,,
\end{equation}
\begin{equation}
\left( g_{,\,nn} - 2\, \frac{f_{,\,n}}{f} g_{,\,n} \right) \left[
g^{2} f^{2} + 2 \,\eta \left ( \kappa f^{2} + g_{,\,n} \,g_{,\,s}
\right) \right] \frac{1}{f^{2} g^{3}} = - \frac{\alpha}{3}\,
(\phi_{,\,n})^{2}\,\, \label{42}
\end{equation}
while,  the dilaton field equation in the lightcone coordinates
takes the form
\begin{equation}
\alpha\,\left[ \phi_{,\,ns} + \frac{3}{2} \left( \phi_{,\,n}\,
\frac{g_{,\,s}}{g} + \phi_{,\,s} \,\frac{g_{,\,n}}{g} \right)
\right] \frac{1}{f^{2}} = -\frac{\beta_{2} \beta
}{2}\,\frac{Q^{2}}{g^{6}}\, e^{-\beta_{2} \phi} - \Lambda\,
\beta_{1}\, e^{- \beta_{1} \phi}\,\,. \label{43}
\end{equation}
Next, we consider the following cases:\\[2mm]
 {(i)}~~~~~~ $ \phi \not = constant \,$,\\[2mm]
the dilaton couples to the other fields. In this case, it follows
from equations (\ref{41}), (\ref{42}) and (\ref{43}) that for $
g_{,\,s} \not =0 $ and $ g_{,\,n} \not = 0 $ the solutions are in
general not static. On the other hand, it may happen that either $
g_{,\,n}= 0 $ or $ g_{,\, s}=0 $ which, in turn, leads to either $
\phi_{,\,n}=0 $ or $ \phi_{,\,s} =0 $. In both cases equation
(\ref{43}) is satisfied only for $ \beta_{1}=0 $ and $ \beta_{2} =
0 $ . However, the general solution is still time-dependent.\\[2mm]
 {(ii)}~~~~~~ $ \phi = constant \, $,\\[2mm]
the dilaton decouples and  may have a constant value. In this case
equations (\ref{41}) and (\ref{42}) reduce to the form
\begin{equation}
\left(g_{,\,ss} - 2\, \frac{f_{,\,s}}{f}\, g_{,\,s} \right) \left[
g^{2} f^{2} + 2 \,\eta ( \kappa f^{2} + g_{,\,n} \,g_{,\,s} )
\right] = 0\,\,, \label{44}
\end{equation}
\begin{equation}
\left( g_{,\,nn} - 2\, \frac{f_{,\,n}}{f}\, g_{,\,n} \right)
\left[ g^{2} f^{2} + 2\, \eta ( \kappa f^{2} + g_{,\,n} \,g_{,\,s}
) \right] = 0 . \label{45}
\end{equation}
Having in these equations either $ g_{,\,n}=0 $ or $ g_{,\,s}=0 \,$,
we get the relation $ g^{2} = - 2 \,\eta \kappa\, $. This results in
two distinct kinds of solutions; the first  ones are flat solutions,
while the second kind of  solutions are time-dependent and given by
\begin{equation}
{\mbox{\boldmath${g}$}} = f^{2} (t,w) ( - dt^{2} + d w^{2} ) + (-2
\,\eta \kappa)\,\, \frac{ d x^{2} + d y^{2} + d z^{2} }{ \left( 1
+ \frac{\kappa\, r^2}{4} \right)^{2} } \,\,, \label{46}
\end{equation}
where $ \kappa \not = 0 $ and the coupling constants must satisfy
the relations
\begin{equation}
\frac{ \beta Q^{2} }{8 \eta^{2} } = - \kappa \,\,,~~~~~~ \Lambda =
\frac{1}{\eta} . \label{47}
\end{equation}
For $ g_{,\,n} \not = 0 $ and $ g_{,\,s} \not = 0 $, it follows
from equations (\ref{44}) and (\ref{45}) that either
\begin{eqnarray}
g_{\,ss} - 2 \,\frac{f_{,\,s}}{f}\, g_{,\,s}&=&
0\,\,,~~~~~g_{,\,nn} - 2 \frac{f_{,\,n}}{f}\, g_{,\,n} = 0
\label{48}
\end{eqnarray}
or
\begin{equation}
g^{2} f^{2} + 2 \,\eta ( \kappa f^{2} + g_{,\,n} \,g_{,\,s} ) =
0\label{49}
\end{equation}
Let us first start with equations in (\ref{48}). They can be
easily integrated to yield the solution
\begin{equation}
f^{2} = N^{\,\prime} g_{,\,s}  = S^{\,\prime}  g_{,\,n}\,\,,
\label{50}
\end{equation}
where $ N= N(n) $ and $ S = S(s) $ are arbitrary functions, the
{\it prime} stands for the total derivative with respect to the
argument and $ g = g (N + S) \,$.  It is clear  that one can make
a new conformal transformation
\begin{equation}
N = \frac{\tilde t + \tilde w}{2} , \qquad   S = \frac{\tilde t-
\tilde w }{2} \label{52}
\end{equation}
that puts the solution in the form $ g = g(\tilde w) $ and thereby
exhibiting its locally static character. Clearly, this proves the
Birkhoff staticity theorem.

Turning now to equation (\ref{49}), which is  equivalent to the
equation  $ 1 + 2 \,\eta A = 0 \,$, and comparing the latter with
equation (\ref{33}) or (\ref{34}) we find that $ g $ must be
constant. This contradicts  our original assumption that $
g_{,\,n} \not = 0 $ and $ g _{,\,s} \not = 0 $. Thus, no
fine-tuning condition exists in this case. This allows us to
deduce that the Birkhoff staticity theorem is valid in the
axi-dilaton gravity model which in addition to the cosmological
constant and Euler-Poincar\'{e} term  also includes a constant
dilaton and an arbitrary axion charge. \\[2mm]
{(iii)}~~~~~~ $ \phi = constant \, $,~~~~~~ $H = 0 $\,\,,\\[2mm]
i.e. the axion charge vanishes. Using equations given in case (ii),
it is easy to check that when either $ g_{,\,n } = 0 $ or $ g_{,\,s}
= 0 $, we have only flat solutions. For $ g_{,\, n} \not = 0 $ and $
g_{,\,s} \not = 0  $ we obtain either static or time-dependent
solutions.  Indeed, as in the previous case discussed  above,
equations in (\ref{48}) admit a locally static solution. That is,
Birkhoff's theorem is true. Meanwhile, equation (\ref{49}) gives
\begin{equation}
f^{2} = \frac{2 \,\eta \left[ (g_{,\,w})^{2} - (g_{,\,t})^{2}
\right]}{2 \kappa\, \eta + g^{2} }\,\,, \label{53}
\end{equation}
where the Euler-Poincar\'{e} coupling parameter $ \eta $ is
related to the cosmological constant by the fine-tuning condition
\begin{equation}
\Lambda = \frac{3}{2 \,\eta }\,\, . \label{54}
\end{equation}
This can be easily verified making use of (\ref{33}) or
(\ref{34}). Finally, the general solution is given by
\begin{equation}
{\mbox{\boldmath${g}$}} = \frac{ 2 \,\eta \left[ (g_{,\,w})^{2} -
(g_{,\,t})^{2} \,\right] }{2 \kappa \,\eta + g^{2} } ( - d t^{2} +
d w^{2} ) + g^{2} (t,w) \, \frac{ d x^{2} + d y^{2} + d z^{2} }{
\left( 1 + \frac{\kappa \,r^2}{4} \right)^{2} } \,\,. \label{55}
\end{equation}
We see that unlike the case $(ii)$, the nonvanishing cosmological
constant obeying the condition (\ref{54}) violates Birkhoff's theorem. The presence of a constant dilaton does not change the situation. That is, the Birkhoff theorem  is valid {\it if and only if} the fine-tuning condition (\ref{54}) is not satisfied.

\section{Static solutions}

In the previous section we have proved the Birkhoff's theorem
without constructing explicit solutions. Namely, we have started
from a general time-dependent metric ansatz (\ref{17}),  for which
the field equations have driven us to a locally static solution.
In this section we give  some explicit examples of static
solutions to our model. It is straightforward to show that in the
static case the field equations (\ref{hc})-(\ref{36}) reduce to
the following equations
\begin{eqnarray}
\frac{1}{g^{2}} \left[ \kappa - \left( \frac{g^{\,\prime}}{u}
\right)^{2} \right]\left[1- \frac{2\,\eta}{u f}\,\left(\frac{
f^{\,\prime} }{u} \right)^{\prime}\,\right] - 2 \,\frac{
f^{\,\prime} g^{\,\prime} }{f g u^{2} } \left[1- \frac{2\,\eta}{u
g}\,\left(\frac{ g^{\,\prime} }{u} \right)^{\prime}\,\right] =
\nonumber \\[1mm]
\frac{1}{u}\left[\frac{1}{f} \left(\frac{f^{\,\prime}}{u}
\right)^{\prime} + \frac{2}{g } \left( \frac{ g^{\,\prime} }{u}
\right)^{\prime}\,\right] +\frac{ \alpha }{2} \left(
\frac{\phi^{\,\prime}}{u} \right)^{2} + \,\frac{\beta Q^{2} }{2
\,g^{6}} \,\,e^{- \beta_{2} \phi} - \Lambda\, e^{ -\beta_{1} \phi }
, \label{56}
\end{eqnarray}
\begin{eqnarray}
\frac{3}{g^{2}} \left[ \kappa - \left(\frac{g^{\,\prime}}{u}
\right)^{2} \right] \left(1-  \frac{2\,\eta }{f
g}\,\frac{f^{\,\prime} g^{\,\prime} }{ u^{2}}\right) - 3
\frac{f^{\,\prime} g^{\,\prime}}{u^{2} f g} = \nonumber \\[1mm]
- \frac{ \alpha }{2} \left( \frac{\phi^{\,\prime}}{u} \right)^{2}
- \Lambda e^{-\beta_{1} \phi} - \frac{\beta Q^{2}}{2 g^{6}}
e^{-\beta_{2} \phi} \,\,, \label{57}
\end{eqnarray}
\begin{eqnarray}
\frac{3}{g^{2}} \left[ \kappa - \left(\frac{g^{\,\prime}}{u}
\right)^{2} \right] \left[1- \frac{2\,\eta}{u g}\,\left(\frac{
g^{\,\prime} }{u} \right)^{\prime}\,\right] - 3
\left(\frac{g^{\,\prime}}{u}\right)^{\prime} \frac{1}{u g} =
\nonumber \\[1mm]
\frac{ \alpha }{2}  \left( \frac{\phi{\,\prime}}{u}
\right)^{2}
- \Lambda \,e^{- \beta_{1} \phi } - \frac{\beta \,Q^{2} }{ 2
\,g^{6} }\, e^{- \beta_{2} \phi } \label{58}\,\,,
\end{eqnarray}

\begin{equation}
\alpha \left( \frac{ \phi^{\,\prime} f g^{3} }{u} \right)^{\prime}
\frac{1}{f g^{3} u} = \frac{\beta_{2} \beta \,Q^{2}}{2 \,g^{6}}\,
e^{- \beta_{2} \phi} + \Lambda \beta_{1} e^{- \beta_{1} \phi}\,\,
. \label{59}
\end{equation}
Here {\it primes} denote the total derivatives with respect to the
spacelike coordinate  $ w  \,$. We consider below some special cases;\\[2mm]
\noindent {\bf Case I:}~~~~~~~ $ H = 0\,\,, $ \\[2mm]
a model with no axion charge. In this case we obtain the static
solution with planar symmetry ($ \kappa =0 $)
\begin{equation}
{\mbox{\boldmath${g}$}} = - w^{2} d t^{2} + \frac{1}{w^{2}}\, d
w^{2} + \frac{1}{w^{\frac{2}{3}}}\, ( d x^{2} + d y^{2} + d z^{2}
)\,\, , \label{60}
\end{equation}
\begin{equation}
\phi ( w ) = \phi_{0} \ln | w | \,\,,\label{61}
\end{equation}
provided that the coupling constant $ \beta_{1} = 0 $, the
Euler-Poincar\'{e} coupling constant $ \eta $,  the cosmological
constant $ \Lambda $  and $\phi_{0}$ satisfy the relations
\begin{eqnarray}
\Lambda &= & \frac{2 \,\eta }{27} \,\,,~~~~~~~  \alpha
\phi_{0}^{2} = \frac{4}{3} \left( \frac{2 \eta }{9} - 1 \right)
\label{63} \,\,.
\end{eqnarray}
We see that the metric components in (\ref{60}) diverge at $ w = 0
$. However, it can be easily verified that this is  a coordinate
singularity. For this purpose, we  evaluate the  curvature scalar
$ {\cal R} $ and the curvature invariant $ R_{ab} \wedge \ast
R^{ab} $ for this metric and show that they are given by the
finite values
\begin{equation}
{\cal R} = - \frac{4}{3} \,\,, \qquad \qquad R_{ab} \wedge \ast
R^{ab} = \frac{228}{81} \ast 1 , \label{63}
\end{equation}
that confirms the statement made above.\\[2mm]
\noindent {\bf Case II:}~~~~~~ $ \phi=constant $, ~~~~~$ H\not = 0\,\,, $ \\[2mm]
\noindent a model with a constant dilaton and non-zero axion
charge. This model admits two types of exact solutions: \\[2mm]
\noindent {(i)} The first type of solutions is given  by the
metric
\begin{equation}
{\mbox{\boldmath${g}$}} = - f^{2} d t^{2} + \frac{1}{f^{2}}\, d
w^{2} + w^{2} \, \frac{ d x^{2} + d y^{2} + d z^{2}}{ \left( 1 +
\frac{\kappa \,r^{2}}{4} \right)^{2} }  \label{64}
\end{equation}
where
\begin{equation}
f^{2} = \kappa + \frac{w^{2}}{2 \eta } \left( 1 \mp \sqrt{ 1 + 2
\eta \left( \frac{2 M}{w^{4}} + \frac{\beta Q^{2}}{3\, w^{6}} -
\frac{ \Lambda }{3} \right) } \right) \label{65}
\end{equation}
and the parameter $ M $ is related to the gravitational mass of
the source. We recognize the well-known static charged black hole
solution with a cosmological constant to Einstein-Gauss-Bonnet
gravity \cite{rong, lidsey} in which the electric charge
is superseded  by an axion charge.\\[2mm]
\noindent (ii) The second type of solutions is given the metric
\begin{equation}
{\mbox{\boldmath${g}$}} = - f^{2} d t^{2} + \frac{1}{f^{2}}\, d
w^{2} + g_{0}^{2} \, \frac{ d x^{2} + d y^{2} + d z^{2} }{ \left(
1 + \frac{\kappa \, r^{2}}{4} \right)^{2} } \,\,,\label{66}
\end{equation}
where $g_{0}$ is a constant and
\begin{equation}
f^{2} =  b\, w^{2} + c_{0}\, w + c_1\,\,. \label{67}
\end{equation}
Here we have introduced the notation
\begin{equation}
b =  \left( 1 + \frac{2 \kappa  \eta }{g_{0}^{2}} \right)^{-1}
\left( \Lambda + \frac{\kappa }{g_{0}^{2}} - \frac{ \beta Q^{2} }{ 2
g_{0}^{6}} \right) \,\,. \label{67'}
\end{equation}
Furthermore, the constants $ g_{0} $, $ Q $ and $ \Lambda $ obey the
relation
\begin{equation}
\frac{3\kappa}{g_{0}^{2}} = - \frac{ \beta Q^{2} }{ 2 g_{0}^{6} }
- \Lambda  \,\,\label{68}
\end{equation}
and $ c_{0} $ and $ c_1 $ are arbitrary integration constants. The
singularity structure of this metric can be understood by studying
its Killing horizons \cite{klosch, daniel}.  It is easy to see
that the metric (\ref{66}) has  coordinate singularities at the
points
\begin{equation}
w_{\mp} = \frac{-c_0 \mp \sqrt{c_{0}^2- 4 b c_1}}{2b}
\,\,\label{68'}
\end{equation}
which are governed by the Killing horizon equation $ f^2=0 $.
However,  evaluating the curvature scalar and curvature invariant
for this metric, we find the constant values
\begin{equation}
{\cal R} = \frac{6 \kappa}{g_{0}^2} - 2 b \,\,, \qquad \qquad R_{ab}
\wedge \ast R^{ab} = \left(\frac{6 \kappa^2}{g_{0}^4} +2 b^2 \right)
\ast 1\,. \label{63'}
\end{equation}
It follows that the singularities at the points (\ref{68'}) are
indeed the coordinate singularities. \\[2mm]
\noindent {\bf Case III:}~~~~~~ $ \phi \not =constant $, ~~~$ H\not = 0\,$,~~~ $ \kappa =0 \,$,~~~ $ \eta =0\,$, \\[2mm]
\noindent a planar model with dilaton and axion fields. This model
admits  a class of solutions with
\begin{equation}
{\mbox{\boldmath${g}$}} = - f^{2} d t^{2} + \frac{1}{f^{2}}\, d
w^{2} + w^{\frac{2}{3}}\, \left(d x^{2} + d y^{2} + d
z^{2}\right)\,\,, \label{69}
\end{equation}

\begin{equation}
\phi(w) = \phi_{0} \ln | w |\,\, , \label{71}
\end{equation}
where
\begin{equation}
f^{2} = \gamma_{0} \,w^{- \beta_{2} \,\phi_{0} } + \gamma_{2}\,
w^{2- \phi_{0} \beta_{1} } + \frac{c_2}{\phi_{0}\, \alpha }
\label{70}
\end{equation}
and
\begin{equation}
\gamma_{0} = - \frac{3 \beta Q^{2} }{4} \,\,, \qquad \gamma_{2} =
\frac{\beta_{1} \Lambda }{\left( 2 \phi_{0} \alpha - \frac{2}{3}
\beta_{1} \right)} \,\,,\label{72}
\end{equation}
 $ c_2 $ is an integration constant and  $\phi_{0} $, $ \beta_{1} $ and
$ \beta_{2} $ must satisfy the relations
\begin{equation}
\phi_{0}^{2} = \frac{2}{3 \alpha} \label{73}
\end{equation}
and
\begin{equation}
\phi_{0} \left( \phi_{0} \beta_{1}^{2} - \frac{10}{3} \beta_{1} +
\frac{1}{2} \phi_{0} \beta_{2}^{2} - \frac{2}{3} \beta_{2} \right)
+ \frac{8}{3} = 0 \,\,. \label{74}
\end{equation}
It should be noted that at least one of the constants $ \beta_{1}
$ and $ \beta_{2} $ is assumed to be non-zero.\footnote{One can
start with ansatz  $ g(w) = w^{s} $ and  obtain a more general set
of  constraint equations where $ s $  varies in the interval $ 0 <
s < 1 $\,. } The curvature scalar and curvature invariant for this
class of solutions are given by the expressions
\begin{equation}
{\cal R} = \gamma_{0} \left[ \frac{2}{3} + \beta_{2} \phi_{0} ( 1
- \beta_{2} \phi_{0} ) \right] w^{-\beta_{2} \phi_{0} - 2 } +\,
\gamma_{2}  \left[ \frac{2}{3} - (2 - \phi_{0} \beta_{1} ) (3 -
\phi_{0} \beta_{1} ) \right] w^{- \phi_{0} \beta_{1} }
\end{equation}
and
\begin{equation}
R_{ab} \wedge \ast R^{ab} = \left( B_{1} w^{-2 \beta_{2} \phi_{0}
-4} + B_{2} w^{-\beta_{2} \phi_{0} - 2 - \phi_{0} \beta_{1} } +
B_{3} w^{- 2 \phi_{0} \beta_{1} } \right) \ast 1\,\,,
\end{equation}
where $ B_{1} $, $B_{2}$ and $B_{3} $ are the constants determined
by the model parameters $\beta_{1}$, $\beta_{2}$, $\phi_{0}$,
$\gamma_{0}$ and $ \gamma_{2} $ . We see that for certain values
of the parameters there are curvature  singularities at $ w
\rightarrow 0 $, whereas for the parameters  obeying the
constraints  $ \beta_{2} \phi_{0} + 2 \leq 0 $  and  $  \phi_{0}
\beta_{1} \leq 0 $ we have a regular solution at this limit. Thus,
for this class of static solutions the singularity structure  is
sensitive to the value of the model parameters and under certain
constraints on their values, there exist asymptotically non-flat
solutions with regular behaviour at $ w \rightarrow 0 $.

\section{Conclusion}

\noindent In this paper we have  studied the exact solutions  to
the effective field equations of an axi-dilaton gravity model
which, in addition to the usual Einstein-Hilbert term with a
cosmological constant, also includes a second order
Euler-Poincar\'{e} term. We have concentrated on five-dimensional
spacetimes  admitting maximally symmetric spatial three-sections
and given a number of interesting  exact solutions. These
solutions basically form two classes. The first class consists of
time-dependent solutions which are valid under certain fine-tuning
relations between the coupling constants of the model and the
cosmological constant. Solutions in the second class  are locally
static that proves validity of  Birkhoff staticity theorem in the
presence of a constant dilaton and an arbitrary axion charge. We
have also given a special class of static solutions, among them a
spherically symmetric black hole solution carrying an axion
charge.

\end{document}